**A critical examination of crop-yield data for vegetables, maize (*Zea mays L.*) and tea (*Camellia sinensis*) for commercialized Sri Lankan biofilm biofertilizers.**

M. W. C. Dharma-wardana[1*✉], Parakrama Waidyanatha[2]
K. A. Renuka[3§], D. Sumith de S. Abeysiriwardena[4]
Buddhi Marambe[5]

[1] National Research Council of Canada, Ottawa, K1A 0R6, Canada,
& Dept. de Physique, Université de Montréal, Montréal, H3C 3J7, Canada.
https://orcid.org/0000-0001-8987-9071
[2] Soil Science society, P. O. Box 10, Peradeniya, 20400, Sri Lanka.
[3] Field Crops Research and Development Institute, Mahaillupallama 50270, Sri Lanka.
[4] Chemical Industries Colombo, Rice Research & Development Unit, Pelwehera, Dambulla, 21100, Sri Lanka.
[5] Dept. of Crop Science, Peradeniya University, Peradeniya, 20400, Sri Lanka.

*✉ Corresponding Author, email: chandre.dharma-wardana@nrc-cnrc.gc.ca

§ Current Address: Fruit Research and Development Institute, Kananwila, Horana, 12400 Sri Lanka.

**Disclosure statement**
The authors declare that they have no conflict of interest, commercial or academic in regard to the topics addressed here or materials and products discussed here. The authors have not received any specific funding from any organization for this work.

**Declaration**.
Artificial Intelligence (AI) Chatbots and associated AI tools have not been used in preparing this work.

**A critical examination of crop-yield data for vegetables, maize (*Zea mays L.*) and tea (*Camellia sinensis*) for commercialized Sri Lankan biofilm biofertilizers.**

**Abstract**.
With increasing global interest in microbial methods for agriculture, the commercialization of biofertilizers in Sri Lanka is of general interest. The use of a biofilm-biofertilizer (BFBF) commercialized in Sri Lanka is claimed to reduce chemical fertilizer (CF) usage by ~50% while boosting harvest by 20-30%. Many countries have explored the potential of biofilm biofertilizers, but have so far found mixed results. Here we review this BFBF commercialized in Sri Lanka and approved for nation-wide use there. We show in detail that the improved yields claimed for this BFBF fall within the uncertainties (error bars) of the harvest. Theoretical models that produce a seemingly reduced CF scenario with an "increase" in harvests, although this is in fact not so, are presented. While BFBF usage seems to improve soil quality in certain respects, the currently available BFBF promoted in Sri Lanka has negligible impact on crop yields. We also briefly consider the potential negative effects of large-scale adoption of microbial methods.

**Section 1 - Introduction**.
Farmers need to maximize harvests from their farms without degrading the soil. When harvested products are removed (without return to the soil, unlike in nature), the soil loses nutrients that have to be replenished regularly. It suffers changes in pH, soil structure and chemistry, as well as changes in populations of soil organisms. An efficient and ecologically non-destructive approach to farming involves *no-till* agriculture (Huggins et al 2008) where the soil and its ecosystem are left least disturbed, while the lost nutrients, pH imbalances etc., are corrected via inputs of mineral fertilizers and other supplements. Though *no-till* alone is not sufficient to improve environmental quality (Daryanto et al 2020), application of no-till or minimal-till agriculture to staples like rice has been addressed (Yang et al 2020, Du et al 2020, Calcante et al 2019). Herbicides are used to achieve weeding without tilling or disturbing the soil (Gianessi 2014).

Nutrient losses and farm-yard runoff occur when farmers pushed by aggressive marketing utilize excessive amounts of agrochemicals (EU green deal 2020, Amarasiri 2015). These can be mitigated by farmer education, adopting methods of precise application of site-specific time-release agrochemicals (Trenkel 1997, Lawrencia et al 2021) and the use of genetically optimized crops. Exceptional yields have been reported from sub-tropical climates, e.g., Egypt (Tantawi 1997), China (Yuan, 1998), with harvests of 17.1 mt/ha for rice. The maximum harvests in tropical climates like Sri Lanka are lower, being about 11.73 mt/ha (Abeysiriwardena and Dhanapala 2021), while the general average is ~4.3 mt/ha for conventional farming and ~2 mt/ha for organic farming. The latter is sometimes presented as an alternative to farming using agrochemicals when small harvests are acceptable.

Organic farmers (Rodale 1947, Steiner 1924) reject modern methods of genetic manipulation and the use of agrochemicals, and accept low harvests by choice. They cannot resort to *no-till* methods as the latter needs agrochemicals. So organic farmers use manual weeding or flooding (Ponnamperuma 1984). Furthermore, embracing composting would enhance GHG emissions (Jiang et al 2011, He et al 2001, Neue 1996). Thus, organic agriculture has sidelined some of the very tools needed for climate mitigation and sustainability.

Microbial fertilizers have been considered as effective alternatives to chemical fertilizers (Dasgupta et al 2021). Even organic farmers may use them. Microbial fertilizers are more robust when presented as biofilms. Biofilms are complex communities of multiple microbial species which are attached to surfaces or physical interfaces found in nature (Costerton et al 1995, Angus and Hirsch, 2013). While attempts to use them in agriculture have met with many challenges, they are being marketed in Sri Lanka as already effective biofilm biofertilizers (Biofilm.lk 2013; Seneviratne and Wijepala 2017). BFBF use is claimed to reinstate sustainability of degraded agro-ecosystems and boost crop yields with reduced use of chemical fertilizers.

However, as established in this study and in an associated study (Dharma-wardana et al 2023b), the available scientific results show that BFBF commercialized in Sri Lanka (referred to here as BFLk) do not meet any of the claims. The use of BFLk does not influence crop yields or supplement the N. P, and K nutrients needed by plants.
.
In this study we critically review the data on BFBF used in Sri Lanka that are available for vegetables, maize (*Zea mays L.*) and tea (*Camellia sinensis*), in the context of claimed boosted crop yields and reduced use of fertilizers. The issues raised are of importance to other countries potentially adopting microbial fertilizers under heightened political concerns for food security (Dharma-wardana 2023a) and climate mitigation (EU green deal 2020). A separate publication (Dharma-wardana et al 2023b) deals with harvests from rice *(Oryza sativa)*.

**Section 2 - Microbial methods for plant nutrition and soil sustenance**
Biofertilizers provide an intermediate approach between organic agriculture and agriculture using only chemical fertilizers, in using agrochemicals as well as microorganisms. This approach, first investigated by Nobbe and Hintner in 1883 (as quoted in Hartmann et al 2008), seeks to maintain soil quality and to provide nutrients using soil microbes. Plant growth-promoting rhizobacteria (PGPR) can potentially enhance plant growth by a variety of mechanisms like phosphate solubilization, siderophore production, biological nitrogen fixation (Mitter et al 2021, Adesemoye and Kloepper 2009), etc.

While microbes can make insoluble phosphates and other substances bioavailable, it can simultaneously release heavy-metal toxins like cadmium, lead, etc., that are fixed in the soil (e.g., as phosphates, carbonates etc.) thus making them appear in the food chain (Dharma-wardana 2018). Furthermore, enhanced microbial activity leads to enhanced release of greenhouse gases (GHG) from the soil, especially if mechanical weeding and tilling are used. Microbial processes can also convert arsenic in the soil to its more toxic bioavailable form sunder anerobic conditions (Upadhyay et al 2020).

The PGPR inoculants contain beneficial bacteria that are primarily applied to plant roots to enhance plant growth and health. They colonize the root surface and produce growth hormones, fix nitrogen, solubilize nutrients, and suppress plant pathogens. Enhanced soil microbial diversity, and its exploitation in Sri Lankan agriculture have been reported by Kulasooriya et al (2017).

Some investigators, e.g., Ye et al (2020) reported successful reduction of CF requirements as well as increases in harvests. Burbano-Figueroa et al (2022) also find that "Commercial inoculants have the potential to duplicate NPK use efficiency by cassava plants while significantly reducing the use of NPK fertilizers". However, Basu et al (2021) stated that the "application of potential biofertilizers that perform well in the laboratory and greenhouse conditions often fails to deliver the expected effects on plant development in field settings".

The objective behind BFBFs is that biofilm formation will create a more suitable environment for biofertilizers to compete with resident organisms and cope with the heterogeneity of biotic and abiotic factors in soil. The BFBFs use microbial communities resident on carrier material and are applied to the soil or plant surface (Mitter et al 2021, Seneviratne et al 2011, Davey et al 2000). They include bacteria, fungi, etc., that form a biofilm, or slimy layer around themselves providing protection and promoting growth. Biofilms have been used for over three decades for targeted applications like community water purification (CAWST 2023), ideally suited for combatting chronic diseases arising from lack of clean drinking water (Hettithanthri et al 2021). In contrast, BFBF usage in agriculture is mostly reported as being in a state of development.

Furthermore, as pointed out by Hermann et al (2013), most biofilm products have been of poor quality. Almost a decade later, Mitter et al (2021) also stated that "additional studies are needed to test BFBFs efficiency at a field scale and to determine optimal processes for a large-scale production and reliable results". One of our objectives is to resolve these contrasting conclusions, and evaluate the strong claims of Sri Lankan BFBF products.

**Section 2A - Use of microbial fertilizers in Sri Lanka.**
In Sri Lanka, the Department of Agriculture (DOA) had examined the inoculant "Nitragin-S" prepared by the NifTAL project, in the early 1980s. While the project showed some success with soybean, it was discontinued, perhaps due to lack of consistent results. Kulasooriya et al (2021) stated that a principal limitation, viz., lack of a low-cost local carrier material (substrate) to provide inoculants to the farmers, was overcome by their group by adopting a carrier made from modified coir dust and a germplasm of local rhizobia for several crop legumes, viz., soybean, mung bean, vegetable bean, groundnut and the forage legume clover. Their isolation, and the characterization of rhizobia from root nodules of local legume crops seem to have followed the method of Somasegeran and Hoben (1991).

However, most tropical legumes are promiscuous and not specific with regard to their rhizobial requirement for nodulation and nitrogen fixation, and the high competitiveness of indigenous soil rhizobia for nodulation. This often makes the introduced rhizobial inocula ineffective (Bent and Chanway, 1998). Whether the BFBF technology can overcome this and provide good harvests is a major question that has to be resolved (DH-web 2021).

The commercialization of Sri Lankan BFLk began in 2010 (Kulasooriya et al 2021, biofilm.lk 2013) receiving much publicity (Indrajith 2017), and the strong support of the National Institute of Fundamental Studies (NIFS), Sri Lanka. The BFLk Biofilm is said to consist of a community of microbes and extracellular polymeric substances secreted by the resident microbes.

Kulasooriya and Magana-arachchi (2016), writing six years after commercialization of BFBF in Sri Lanka, voiced concerns about microbial methods:

1. They say (page 120): "this low-cost technology ... has been successful only in a few areas in India ...., the yield increases recorded in field experiments ... ranged from 12 to 19 %, which were not higher than the previously reported values.
2. Cyanobacterial biofertilizer technology in Sri Lanka was successful only up to the stage of pot experiments. Most of the cyanobacterial inoculants added to the rice fields could not overcome consumption by the rice field micro-fauna and ...
3. More recently it has been reported that $N_2$ fixing species of cyanobacteria could be incorporated with certain eubacteria and fungi to form biofilm-biofertilizers and preliminary field trials have given *encouraging* results with rice (Kulasooriya & Seneviratne, 2013).

The field trials and research studies using BFBF have continued (Seneviratne et al 2013, Kulasooriya et al 2017, Premarathna et al 2021) using the patented material BFLk. The "farmer acceptance" of the material has been used to "validate" the product. Kulasooriya et al (2021) justify this procedure as follows:

"It had been our practice to conduct field tests … in farmer's fields for two reasons. These fields mimic the … small holder farmers more closely than those in research stations …. Moreover, if the tests give positive results, acceptance of the technology by the farmers is that much easier. … Two well designed field trials with replicates were conducted to obtain N-fertilizer yield response curves for soybean and vegetable bean. The effects of the corresponding rhizobial inoculants were read off from these curves".

Unfortunately, the crop yield response curves, results from controls etc., are not available in the public domain or in peer-reviewed publications.

As the tests have been conducted by individuals with vested interests in the patented product, a clear conflict of interest exists, and an independent review is needed for scientific acceptance. Therefore, although BFLk has been commercialized, doubts have been raised at various times during the last decade (Waidyanatha 2017, DH-web 2021).

In this study we analyze the literature available in the public domain, and conclude that the BFLk claims remain unsubstantiated. The available yield data show that BFLk when used with or without CF has no clear positive effect on harvests, and confirm the difficulties noted by other workers elsewhere (Mitter 2021, Basu et al 2021, Hermann et al 2013) in implementing biofilm biofertilizers. This discussion is presented in the hope that more complete crop-yield data from Sri Lankan BFBF formulations be made available in the public domain, to help agriculturalists to make informed-decisions.

**Section 3 - Crop response to fertilizer inputs and BFBF inputs.**
We discuss the public-domain data as well as field data available through Sri Lanka's Department of Agriculture (DOA).
1. The initial release of BFBF seems to have been justified by a publication in the Commonwealth Agricultural Bulletin (CAB) journal, by Buddhika et al (2016), as well as the publications by Seneviratne et al (2008, 2011, and 2013). Later, A newspaper article (Indrajith 2017) made strong public claims following the CAB publication. In re-examining the article published in CAB, most claims for BFLk seem untenable..
2. Results of field trials on maize (*Zea mays L.*) conducted at the Field Crops Research and Development Institute, Mahaillippallama, Dept. of Agriculture, Sri Lanka (Renuka 2012) are examined and we find that BFLk had no impact on maize harvests.

3. The crop response to BFLk is said to be null at least until the CF fraction is close to 50% (Amarathunga et al. 2018, Wickramasinghe et al. 2018). Usually only the crop yield at 50% CF+BFBF is given and compared with the 100% CF datum; complete yield curves have not been published. Only short abstracts at conferences are available.
4. Studies on rice *(Oryza sativa)* are presented in a separate study (Dhrma-wardana et al 2023b); we summarize the relevant results in the paragraph given below.
   Studies of BFLk by the DOA in 2017 and 2018, hitherto not available in the open literature, provide valuable data on rice harvests with and without BFBF. The Principal Agriculturist of the Rice Research Development Institute of the DOA had advised in December 2020 that more pilot scale trials should be conducted before recommending BFBF for farmer's use (Thilakasiri 2020). The DOA Principal Soil Scientist had advised that "*because of significant yield reduction*", substitution of even 35% of the recommended CF with BFBF is not advisable (Rathnayake 2020). However, the government seems to have approved BFLk usage, with the CF inputs reduced by 20% (DOA 2020). Nevertheless, BFLk commercials recommend reducing CF by 50% and including BFBF for rice cultivation, promising 20-30% boosted harvests. These claims are not supported by the available data.
5. BFBF-T specific for tea has been commercialized. However, only the work of DeSilva (2014), and a one-page abstract (Chandralal 2020) are available as scientific evidence. Chandralal et al (2020) give one data point, but not for 50% CF+BFBF, but for 75% CF+BFBF, with a claimed 16% boost in crop yield. This "boost" falls on the margin of the known uncertainty in *made-tea* yields. The Tea Research Institute (TRI 2021) advisory documents do not mention the use of currently available BFLk.

These will be dealt in greater detail below.

**Section 3.1 - The CAB Journal (2016) publication with erroneous claims for BFBF.**
Buddhika et al (2016), writing in the CAB journal state the following:
"BFBFs have been tested successfully for their fertilizing potential of many crops, such as maize, rice, a wide range of vegetables and for plantation crops like tea and rubber, under greenhouse and field conditions. Their effectiveness under field conditions has made it possible to reduce the use of chemical fertilizer (CF) NPK by 50%, with several other beneficial functions needed for sustainability of the agroecosystems".

Their essential results were presented in their Table 6.1, and *said to substantiate their claims of being able to obtain yields comparable to 100% CF using only 50% CF together with the recommended amount of BFBF*. The earliest misgivings had been voiced by Sarath Amarasiri (2017), a past Director General of the DOA. Concerns were also expressed by Waidyanatha (2017) in response to the Island newspaper article on biofilm-biofertilizers by Indrajith (2017).

The crop yields reported in the CAB journal (Buddhika et al 2016) are reexamined, by comparing with accepted harvests for 100% CF, Dept. of Census and Statistics (DCS 2022), Sri Lanka. Column 4 of Table 1 below shows that the CAB data (columns 2,3) are shockingly outside the domain of any statistical validity by orders of magnitude, in essentially all cases except for rice, maize and tea that are discussed further. For instance, the yields for cabbage in columns 2, 3 are a factor of 40 to 50 times less (i.e., < 2.5%) compared to normally expected yields (column 4)! There is an extreme decrease in harvests on using BFLk.

*Table 1. Columns 1-3: Mean crop yields following application of biofilm- biofertilizer (BFBF) combined with 50% of the recommended rate of chemical fertilizer (50% CF) compared with application of the recommended rate of chemical fertilizer (100% CF) in field experiments conducted in different agroecological regions of Sri Lanka as given in Table 6.1, Buddhika et al (2016) in the CAB journal. Column 4: Data from the Dept. of Agriculture (DOA) and the Dept. of Census and Statistics (DCS) of Sri Lanka for 100% CF application.*

| Crop | Yield (kg/ha) (CABJ) 50% CF+BFBF | Yield (kg/ha) (CABJ) 100% CF | Average and maxi yield (DCS and DOA https://doa.gov.lk/hordi-crops/) 100% CF |
|---|---|---|---|
| **Tea** (*Camellia sinensis*) | 4300 ± 606 | 4100 ± 678 | 3,500, 6,000 |
| **Rice** (*Oriza sativa*) | 4420 ± 715 | 3580 ± 1295 | 4,747  6,622 |
| **Maize** (*Zea mays*) | 2681 ± 322 | 2502 ± 338 | 1,000  5,000 |
| **Radish** (*Raphanus sativus*) | 1192 ± 251 | 992 ± 188 | 20,000  50,000 |
| **Cabbage** (*Brassica oleracea*) | 1302 ± 342 | 980 ± 249 | 40,000  50,000 |
| **Bitter Gourd** (*Momordica charantia*) | 1547 ± 445 | 1563 ± 440 | 20,000  30,000 |
| **Aubergine** (*Solanum melongena*) | 748 ± 175 | 678 ± 260 | 18,000  40,000 |
| **Okra** (*Abelmoschus esculentus*) | 3107 ± 1719 | 1739 ± 710 | 15,000  30,000 |
| **Chilli** (*Capsicum angulosum*) | 3478 ± 1754 | 2350 ± 919 | 25,000   32000 |
| **Hung. Wax pepper** (*Capsicum annuum var annuum*) | 238 ± 50 | 152 ± 39 | 10,000  25,000 |
| **Tomato** (*Solanum lycopersicum*) | 335 ± 86 | 397 ± 131 | 20,000  30,000 |
| **Pole Bean** (*Phaseolus vulgaris*) | 2762 ± 886 | 2396 ± 753 | 9,000,  12,000 |

.

## Section 3.2 - Field trials with BFBF for Maize cultivation.

We find the earliest claims of increased yields with reduced CF application for Maize (*Zea Mays* L.) when supplemented with BFLK in the work of Buddhika et al (2012), although without detailed yield data. However, detailed data on the response of maize to CF+BFBF with the CF component varied from zero to 100% were available from the Mahailluppallama (MI)-DOA Research station, from the work done by one of the present authors (Renuka 2012). The data are displayed in Figure 1.

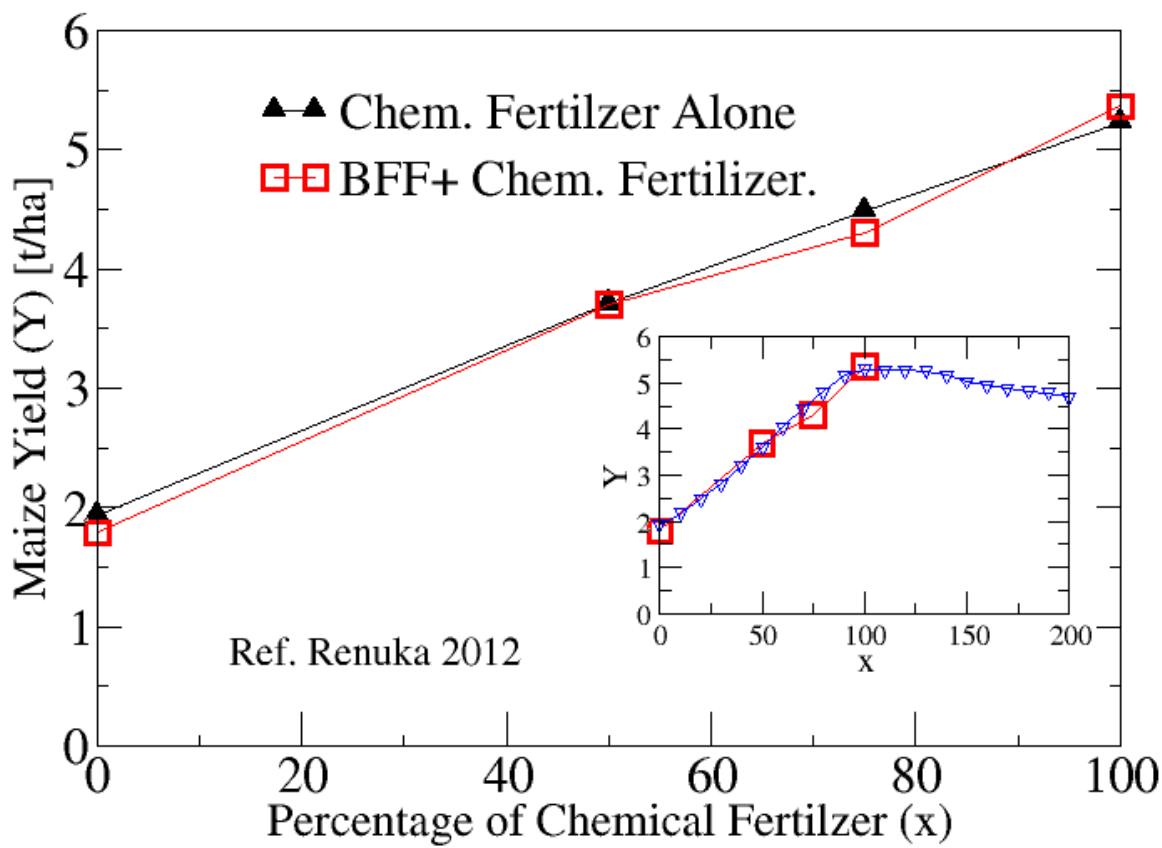

*Figure 1 Maize yield function Y(x) with and without BFBF.*

The main graph displays the crop yield (black triangles) increasing almost linearly with the CF input with a harvest of about 5 mt/ha at 100% CF. The (red) boxes show the harvest when the CF input is supplemented with BFBF. ***The addition of BFBF has no impact on the yield***.

The inset gives an extended yield function (triangles pointing down) *modeled* to include the effect of excess use of fertilizer, beyond the recommended value which is taken as 100% CF. The yield does NOT increase with excess fertilizer. The yield *Y* remains nearly flat beyond the optimal value (100%), and usually curves downwards with increased fertilizer, as shown in the inset plot. The MI-DOA field trial remained within the linear part of the curve.

According to Chathurika et al (2015), an optimal CF for maize would consist of Urea 325, TSP 100, MOP 50, all given in kg/ha, i.e., a total CF of 475 kg/ha. While the DOA trial did NOT enter the excess-fertilizer regime as seen from the graph, we consider the hypothetical case of a trial with 100% CF taken at a significantly higher value, when the yield would be smaller than at the optimal value of CF at 475 kg/ha. In this case, using a smaller amount of CF and BFBF would give a better harvest, and one may erroneously claim that BFBF can successfully reduce the amount of CF needed, and at the same time increase yields. That is, unless the site-specific optimal CF 100% input is known, field trials conducted in the "beyond-the-maximum asymptotic region" of the yield curve can provide wrong conclusions about the efficacy of BFBF in lowering fertilizer inputs and in seemingly boosting harvests above the 100% CF.

This seems to be the case in Premarathne et al (2021) as well, in their study of BFLK for rice use 450 kg/ha of CF when the DOA recommended values is in the 225-300 kg/ha, paving the way for misleading interpretations, as discussed elsewhere (Dharma-wardana et al 2023b).

**Section 3.4 - Use of BFBF in tea cultivation**

A conference on *Biotechnology in agriculture* hosted by the 'Coordinating Secretariat for Science Technology and Innovation' (COSTI 2014), Sri Lanka, contains a contribution entitled *Biofilm biofertilizers, a success story*. It alludes to successful use of BFBF+50% CF to replace 100% CF in trials done in 2005 at the TRI, using 50% T65; trials done in 2008 and 2013 at Ratnapura, using 50% TRI3055, and refers to many farmers' testimonies.

However, the TRI recommendations (TRI 2021) circular 04/2021 or earlier technical documents do not even mention biofilm biofertilizers. This is true even for its most recent publications, although BFLk was tested for potential applications for tea cultivation (DeSilva et al 2014). The circular 04/2021 (TRI 2021) specifically states the following:

"Do not curtail fertilizer inputs for tea nurseries and immature tea (up to formative pruning) in the use of T 65, and T 200 and T 750 mixtures respectively and continue as recommended (Refer Advisory Circulars SP1 and SP2)".

In contrast, the BFLK advertisements (Biofilm-T 2014) recommend the use of 50% of T65 chemical fertilizer supplemented with BFLK for tea nurseries, alluding to supporting research at the National Institute of Fundamental Studies (NIFS), Sri Lanka. Unfortunately, yield functions $Y(x)$, i.e., crop-yield versus the amount of CF input, or any pertinent data from field trials since 2005 to date (2023) have not been reported in the peer-reviewed literature, nor in any public-domain records of the annual reports of the NIFS, Sri Lanka.

However, the work of DeSilva et al (2014) provides some results that we summarize below. Trials at two low-elevation experimental stations (Ratnapura nad Kottawa), a mid-elevation station at Elkaduwa, and a high-elevation station at Talawakelle have been reported. Unfortunately, only result at 100% CF, 50% CF, and 50%CF+BFBF have been reported rather than full crop-yield curves in each case. What is encouraging is that there is a slight trend observable where the total soil N, P, K, as well as the microbial biomass, soil organic carbon etc., were higher for the 50%CF+BFBF applications. The crop yield (reported as *made tea* output per hectare) shows an improvement, as summarized in Table 2 and can be considered encouraging *a priori*.

Nevertheless, the gain is well within the crop-yield error bars for tea cultivation even for adjacent tea plots or plantations. As no error bars had been given, we have constructed an estimate applicable for the Talawakelle area for 2014 using crop yield data for seven tea estates in the region, viz., Bearwell, Great Western, Holyrod, Logie, Mattekelle, Palmerstone and Wattegoda estates (Talawakelle PLC 2014). The uncertainty of ± 450 kg/ha is likely to hold for other regions was well. The claimed changes on using BFBF are well within these yield-uncertainty limits.

*Table 2. 2<sup>nd</sup> Year made-tea yield after harvesting commenced, as read off approximately from Figure 11 of DeSilva et al (2014), reporting results of field trials at four locations to test the efficacy of BFBF inoculants in tea cultivation. No error bars have been provided by the authors. We have added the ± 450 kg/ha error bar for the Talawakelle region using data available from the 2014 Annual report of the Talawakelle PLC (2014).*

| Test site | 100% CF kg/ha | 50% CF kg/ha | 50% CF+BFBF kg/ha |
|---|---|---|---|
| **Ratnapura** | 2325 | 1300 | 2300 |
| **Kottawa** | 2400 | 2800 | 2400 |
| **Elkaduwa** | 2300 | 1500 | 2250 |
| **Talwakelle** | 1750 ± 450 | 1400 | 1600 |

In most crops, the non-linear asymptotic region beyond 100% CF either flattens out or becomes a decreasing function of the CF input. In contrast, with tea, because tea leaves are removed in weekly plucking rounds, the tea bushes remain responsive to continued fertilizer-input increases; consequently, no asymptotic flattening or decrease in the yield function $Y(x)$, where $x$ is the CF input, is observed for tea even for large values of $x$ (Owuor 1997, DeCosta et al 2007, Cheruiyot et al. 2009).

Hence using a higher CF input and doing a throwback to 50% cannot be used to create the illusion of a boosted yield with 50% CF+BFBF. We examine the tea crop yield as a function of the CF input to relate it to claims made for the BFBF procedure. However, what are available as results for the BFLk procedure are a few one-page poster abstracts of conference presentations that have not been followed up by full papers in peer-reviewed journals.

Due to the lack of any better material, we consider the conference poster by Chandralal et al (2020). While BFBF advocates claim a 50% reduction in the CF input and a 25% boost in yields, Chandralal et al (2020) claim only a 25% reduction and a boost in the yield of 16%. Only one data point (value at 75% CF input) has been presented.

The study consisted of two uniformly managed tea lands in Badulla, Sri Lanka using as chemical-fertilizer mixture meant for vegetatively propagated mature tea grown in the soils of the Uva province, identified by them as VP/Uva925.

According to Chandralal et al (2020), the fields were applied with two treatments separately:

(a) 100% CF of Tea Research Institute (TRI) recommendation of VP/Uva925
(b) 75% CF of TRI recommendation of VP/Uva925 + BFBF 2.5 L per hectare.

However, there is no TRI recommended fertilizer named VP/Uva925, although there exists a VP/Uva945. So, we assume that the field trials were conducted using VP/Uva945.

They report a yield of 1154±40 kg/ha of made-tea from (b), while (a) with 100% CF gave a yield of 992±33 kg/ha of made-tea. These two data points are displayed in Fig. 5 together with upper and lower bounds of the crop-yield functions which are constructed using the data from TRI publications etc., as discussed below.

The TRI advisory circular TRI 2000-SP3 (TRI publications 2021) describes the fertilizer VP/Uva945 as 28.6% N, 3.8% $P_2O_5$, 14.8% $K_2O$, made up of 587 parts of Urea, 125 parts of Eppawala Rock Phosphate, and 233 parts of MOP, making up a total of 945 parts.

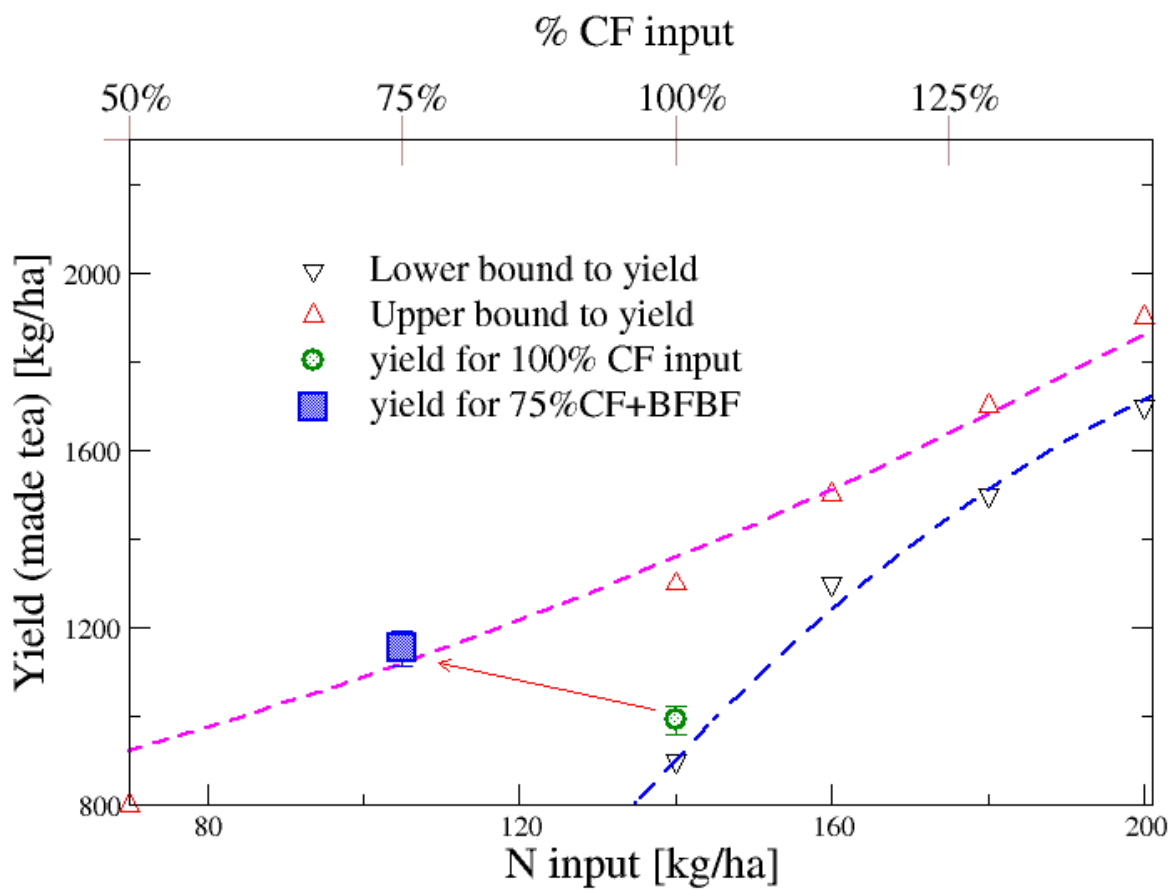

*Figure 2 Crop-yield of tea as a function of CF input for vegetatively propagated (VP) Uva mature tea. The upper and lower dashed curves define the upper and lower bounds to the yield Y(x) that can be expected for a given input x of N or equivalent fertilizer amount per kg/ha. Thus, for 100% CF input (140 kg/ha of N) the crop yield (given as made tea) falls between 900-1300 kg/ha. The triangles are data points from TRI (2000), suitable for VP/Uva945 CF applications. The two data points linked by an arrow are: Chandralal et al (2020), results for 100% CF yield (hatched circle); 75% CF+BFBF (hatched box).*

Reviewing Tables 2 and 3 of the document TRI 2000 (TRI publications 2021), a crop yield 992 kg/ha of made tea with 100% CF corresponds to supplying an amount of fertilizer with 140 kg of Nitrogen. That is, a 100% CF input implies that 490 kg/ha of VP/Uva945 are applied for mature tea grown in the Uva province to obtain a made-tea yield of 900-1300 kg/ha. If higher yields are sought, then higher amounts of CF have to be applied, as seen in the crop-yield curve (Fig. 2) which incorporates data from TRI 2000 SP3 in the range of inputs of 140-300 kg of *N* per hectare. Each *N*-input amount is proportional to the CF input amount, and defines the upper and lower bounds of the crop-yield curve.

Fig. 2 shows that the BFBF+75% CF procedure falls marginally on the upper bound of the expected crop uncertainty, and provides no evidence in favour of the BFBF claim that it can lower CF usage to 50% and boost the yield by ca 25%. In fact, no data at $x = 50$% CF +BFBF, and at any other values of $x$ + BFBF are given by BFLk advocates.

**Section 3.5 – Behaviour of BFBF at lower CF fractions**
Premarathna et al (2021) state that Amarathunga et al. (2018) and Wickramasinghe et al. (2018) showed that the application of BFBF alone could not support an improved plant growth in rice. Normally, when the soil N-content is high, N-fixing organisms become less active. So, one would expect BFBF to work well at *lower* CF loadings, and *not* for higher CF loading. This has in fact been observed in assessing the NPK-use efficiency of commercial inoculants available in Colombia for cassava (*Manihot esculenta* Cratz) where Burbano-Fugueroa et al (2022) used a sophisticated data-development-analysis methodology.

The yield of maize obtained with CF and CF+BFBF in the form of BFLK led us to conclude that BFLk has a null effect on the harvest. The use of the non-linear part of the yield curve in CF+BFBF trials would explain why BFBF does not work alone or with lower fractions of CF. The advocates of BFLk claim that 50% CF should be used with BFLk to best observe its effect. In the case of maize; this corresponds to 162.5 kg/ha of urea, and BFLk is evidently unable to supply any N by its own action.

If the application of BFLk improves the soil-plant-microbial interactions and leads to better nutrient use and increased yield while cutting down the CFs (Seneviratne et al., 2008a; Premarathna et al., 2018) then the failure of BFBF to act at low CFs is intriguing. We interpret this to mean that the development of strong microbial colonies in the soil triggered by BFLk occurs (if at all) only at high CF inputs, and is somewhat analogous to the rise of toxic algal blooms when nutrient concentrations become sufficiently large in aquatic systems. This suggests that if the onset of any BFBF action is circa 50% CF or higher, this would also be associated with a steep rise in $CO_2$, $N_2O$ and other GHG emissions (Baggs 2011, Sabba 2018). This aspect of biofertilizer usage has received scant attention.

**Section 4 - Discussion.**
While the concept of using microbes to provide at least some of the fertilizers needed for agriculture is attractive, any attempt to assume that this target has been achieved is fraught with difficulty because crop yields as well as soil quality are determined by the dynamics of a complex ecosystem, as represented in Fig. 1 of Dharma-wardana et al (2023b).

The commercialization of BFBF in Sri Lanka circa 2010 seems to have been driven by the political demands of the day, as reflected in newspaper articles, e.g., by Indrajith (2017), BFBF websites (Biofilm.lk 2013) and press interviews (Piyatilleke 2015). Nevertheless, even in 2016 scientists who were themselves associated with the product would, in their academic reviews, state that only pot-experiments exist to show the potential of these products.

The advocates of BFLk usage claim that they have hundreds of results from field trials with farmers, as well as with Department of Agriculture scientists, but those in the public domain do not substantiate their claims. The seeming reliance on a single data point or two without any publication of complete crop-yield functions casts doubt on whether sufficient sets of field data have been used.

The use of urea, TSP etc., that have not been formulated as time-release products, and the imprudent use of CF certainly lead to problems of nitrogen loss and environmental pollution. On the other hand, increased use of microbial methods correspondingly increases the output of greenhouse gases consequent to increased soil microbial activity, just as with the microbial decomposition of farm-yard waste into organic manure.  Enhanced emission of $CO_2$, $CH_4$ and oxides of nitrogen results.

Furthermore, forcing large increases of a few varieties of microbial populations in the soil supported by 50% CF and using externally supplied inoculants, if practiced on a large scale may lead to unexpected long-term consequences that are currently not understood. The process is analogous to artificially creating the analogue of algal blooms within the soil. When chemical fertilizers are used irrationally, their wash off creates agricultural pollution. Similar negative results can occur with soil algal blooms.

With 50% CF and large microbial populations triggered by biofertilizers, the run-off will consist of microbial biomass rich in nutrients, together with microbial forms evolving in different directions in waterways and outside the agricultural environment.  Unlike inorganic processes, microorganisms can mutate and respond along unexpected pathways and due caution is very important.

While writers proposing various alternative agricultural strategies are quick to claim that agricultural soils have deteriorated due to the use of chemical fertilizer, they fail to note that even soils farmed without use of fertilizers deteriorate; they have to be left fallow (DeSilva et al 1991) or properly managed to retain soil quality. As many experienced tea planters have also noted (e.g., Talawakelle PLC 2014), well known practices of using cover crops and soil management can sustain the quality of soils. Unfortunately, the plantations sector in Sri Lanka suffered enormously due to the deficit of know-how and management created by the process of nationalization and re-privatization. Other countries embarking on such projects need to learn from the Sri Lankan experience and observe due caution.

As the concepts behind the BFBF procedure have scientific merit, the method should be further studied; Sri Lanka's NIFS that has prematurely and imprudently backed the product must rectify its past failure in stewardship by setting up a robust publicly available  data base for BFBF practice obtained from experiments by scientists with no competing commercial interests.

**Section 5. Conclusion.**
We conclude that the publicly available data for crop yields when BFLk is used with chemical fertilizers can be well explained by assuming that the BFBF addition *contributes nothi*ng to the crop yield. Published data on vegetables, maize, tea have been examined in detail and the claims are found to be unsubstantiated.  In most cases the 50% reduction in CF used in the BFBF practice actually brings the CF input closer to the inputs recommended by the department of agriculture of Sri Lanka, and moving away from the excessive inputs of farmers. The disappointing results obtained so far from BFLk should be taken as a spring board for further research into developing successful biofilm biofertilizers for agricultural use.

.

**References.**